\documentclass[aps,preprint,groupaddress]{revtex4}

\usepackage{natbib}
\usepackage[dvips]{graphicx}
\usepackage{amsmath}
\usepackage{amssymb}

\begin{document}

\title{Simulation and assessment of ion kinetic effects in a direct-drive capsule implosion experiment}

\author{A. Le}
\author{T. J. T. Kwan}
\author{M. J. Schmitt}
\author{H. W. Herrmann}
\author{S. H. Batha}
\affiliation{Los Alamos National Laboratory, Los Alamos, New Mexico 87545, USA}

\begin{abstract}
The first simulations employing a kinetic treatment of both fuel and shell ions to model inertial confinement fusion experiments are presented, including results showing the importance of kinetic physics processes in altering fusion burn. A pair of direct drive capsule implosions performed at the OMEGA facility with two different gas fills of deuterium, tritium, and helium-3 are analyzed. During implosion shock convergence, highly non-Maxwellian ion velocity distributions and separations in the density and temperature amongst the ion species are observed. Diffusion of fuel into the capsule shell is identified as a principal process that degrades fusion burn performance.
\end{abstract}

\maketitle
\section{Introduction}
The design and modeling of inertial fusion capsules rely on hydrodynamic codes. In the hydrodynamic (or fluid) limit, the mean-free path of particles is assumed small compared to gradient length scales and the particle velocity distribution is nearly Maxwellian. These conditions are not always satisfied in experiments, and a kinetic description that tracks the particle distribution in phase space is more appropriate. Ion kinetic effects outside the scope of single-component hydrodynamics have been proposed to explain fusion yield anomalies in recent experiments using glass capsules filled with a mix of deuterium, tritium, and helium-3 fuel at the OMEGA laser facility of the Laboratory for Laser Energetics (LLE) and at the National Ignition Facility (NIF) \cite{rygg:2006,herrmann:2009,casey:2012,rosenberg:2014,rinderknecht:2015}. While theoretical investigations \cite{amendt:2010,amendt:2011,amendt:2015} and multi-fluid numerical studies \cite{bellei:2013,bellei:2014} may guide our understanding of kinetic effects in fusion experiments, they fall short of offering a self-consistent model applicable over a broad range of regimes. Meanwhile, kinetic treatment via Vlasov-Fokker-Planck methods \cite{larroche:2012,larroche:2016} has quantified ion species separation, non-Maxwellian particle distributions, and other kinetic effects. It was not possible, however, to include the capsule shell dynamics in these previous kinetic studies, leaving large uncertainties compared to experimental observations. 

Here, we use particle-in-cell (PIC) kinetic techniques to model a pair of direct-drive capsule implosion experiments performed at the OMEGA facility \cite{herrmann:2009}. These capsules were in an early stage of compression at the end of the laser pulse. Single-component fluid codes such as HYDRA \cite{marinak:1998} can adequately model the ablation process that accelerates the pusher. In order to simulate the remaining convergence and stagnation phase of the implosion with a kinetic treatment of the ions, the conditions from the HYDRA simulation at the end of the laser pulse are linked to the PIC code LSP \cite{welch:2001}. Our PIC simulations include a kinetic treatment of both the fuel and shell ions, are in reasonable quantitative agreement with experimental fusion burn measurements, and allow an assessment of the importance of various kinetic and multi-species effects in fusion burn performance.

The paper is outlined as follows. We describe the simulation set-up and the experiments modeled in the next section. This is followed by a description of the kinetic simulation results including an assessment of the importance of kinetic effects for fusion burn, and a summary discussion concludes.

\section{Simulation Set-up}  
In the considered experiments, glass (SiO$_2$) shells of thickness $4.7$ $\mu$m and a diameter of 1098 $\mu$m were filled to a pressure of 5.1 atm with a 50/50 equimolar mix of deuterium/tritium (D/T) gas. Three such capsules were used in the first set of experiments. In the second set, three capsules were filled with an additional 5.6 atm of $^3$He gas, yielding a molar mix of 32/32/36 D/T/$^3$He. Note that these implosions are not "hydro-equivalent" \cite{rygg:2006}. In all cases, the target implosion was driven directly with 16.3 kJ total laser energy in a 0.6 ns square pulse from 60 beams of 351 nm UV light. 

Ion kinetic effects are modeled using LSP, a PIC code with fluid or kinetic options for each plasma species \cite{welch:2001,welch:2006,thoma:2011}. Because we focus on ion time scales, the electrons are treated as a fluid. An electron heat flux limiter of $0.06$ is chosen to match the corresponding radiation-hydrodynamic simulations, although it minimally effects the LSP results. The kinetic ion equations of motion are solved by standard PIC methods. Each ion species is treated separately, except that the SiO$_2$ glass shell is modeled by a mean ion with $Z_S=10$ and $m_S/m_p=20$ ($m_p$ is the proton mass). This is appropriate because the shell material is much more highly collisional than the gas, and minimal separation of the Si and O ions is expected. The inter- and intra-species ion collision operators are computed with the Nanbu \cite{nanbu:1997} formulation of the Takizuka-Abe \cite{takizuka:1977} particle-pairing algorithm. We use a direct implicit algorithm \cite{hewett:1987,smithe:2009} to relax electric field evolution time-step constraints and allow a reasonable turnaround time. The simulations are performed in one-dimensional spherical geometry with reflecting boundary conditions  at both ends and 2000 cells covering the range from $r=3\times10^{-4}$ cm to $r=0.1$ cm. The origin is excluded to avoid numerical instability, and the outside boundary is located sufficiently far to have negligible influence on the capsule. At least a few thousand particles per cell are necessary for numerical convergence and to reduce noise near the origin, and we initialize our runs with 5000 numerical particles per ion species per cell. Exact energy conservation is not maintained by this algorithm, and numerical cooling results in a loss of $\sim 5\%$ of the total energy by bang time. The typical chosen time step of $dt\sim1.7$ fs resolves all ion collision frequencies except the shell self-collisions in cooler regions, where $\nu_{SS}\times dt \sim$1 to 10.

\begin{figure}
\includegraphics[width = 10cm]{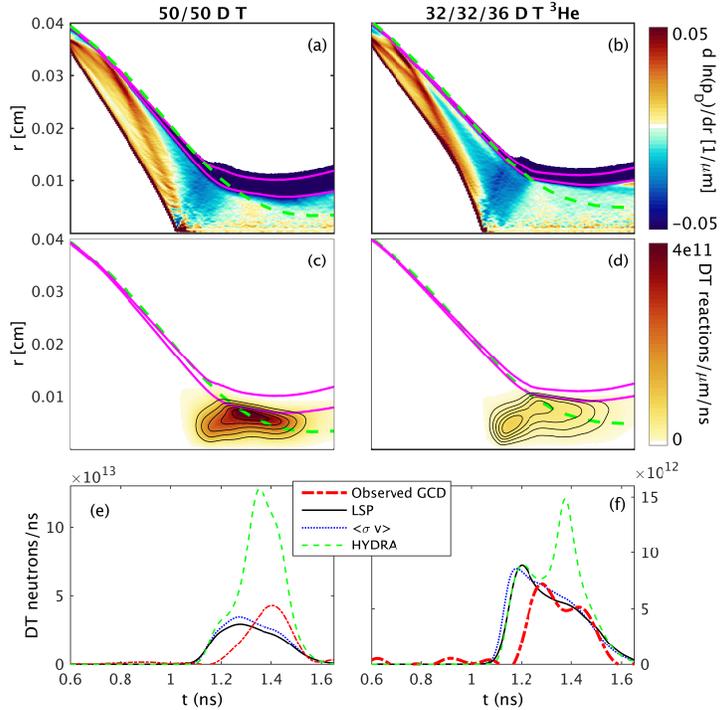}
\caption{(a-b) Inverse gradient length scale $d \log(p_D)/dr$ of the D pressure profile in an $r-t$ diagram. Magenta contours show the fuel-shell interface at locations of $10\%$ and $90\%$ SiO$_2$ molar concentration. The green dashed curve is the fuel-shell interface from the HYDRA calculation. (b-c) DT fusion reaction rate per micron shell from temperature-based mean $<\sigma v>$ in the kinetic calculation.  (e-f) Burn histories measured by the GCD diagnostic (red dash-dotted), inferred from the HYDRA calculation (green dashed), and computed from LSP with a Monte Carlo fusion package (black) and based on $<\sigma v>$ (blue dotted). Simulation burn histories are convolved with a Gaussian of variance $\sim35$ ps to mimic GCD instrumental response.
\label{fig:burn}}
\end{figure}

The initial conditions are extracted from a 1D radiation-hydrodynamic HYDRA \cite{marinak:2001} simulation at time $t=0.6$ ns at the end of the laser pulse. At this point, the converging shock has reached a radial position of $\sim350$ $\mu$m in the fuel. Based on the bulk density, velocity, and temperature profiles, the kinetic ion species are initialized with drifting Maxwellian velocity distributions. The remaining "coasting" phase is simulated in LSP out to $t\sim1.6$ ns, which includes shock convergence and the compression phase of fusion burn. These experiments were selected in part because the coasting phase is relatively long, which allows the incoming shock solution to relax to the kinetic PIC physics and leaves time for fuel species stratification to develop. A series of HYDRA simulations were carried out with varying input laser energy, which was scaled by factors of 0.76 to 0.9 to account for backscatter (not measured at the time of this experiment) and 1D laser propagation effects. The total DT yield from each run is listed in Table 1. We focus below on the runs with a factor of 0.85.

\begin{table}
\begin{ruledtabular}
\begin{tabular}{c|c|c|c|c}
 & \multicolumn{2}{c|} {D/T} & \multicolumn{2}{c}{D/T/He3} \\
 & \multicolumn{2}{c|} {(Yield $8.9\times10^{12}$)} & \multicolumn{2}{c}{(Yield $1.6\times10^{12}$)} \\
\hline
Laser Factor & HYDRA 	& LSP 	& HYDRA 	& LSP \\
\hline
0.76         & 0.65	& 1.56  & 0.94		& 1.21\\
0.8          & 0.49	& 1.31	& 0.70		& 0.85\\
0.85         & 0.34	& 0.96	& 0.46		& 0.64 \\
0.9          & 0.24	& 0.72	& 0.30		& 0.47
\end{tabular}
\end{ruledtabular}
\label{tab:yoc}
\caption{Observed DT neutron yield divided by simulated yield for varying laser factors. The total laser energy injected into the initial rad-hydro simulation was scaled by the laser factor to account for the (unmeasured) backscatter and 1D laser propagation effects.}
\end{table}

\section{Hybrid Kinetic Simulation Results}

An overview of the implosion dynamics from the LSP kinetic simulations is in Figs.~\ref{fig:burn}(a-b), which shows the inverse gradient length scale $d \log (p_D)/dr$ of the D pressure in an $r-t$ diagram. The incoming shock converges at $t\sim 1.1$ ns, and a weaker rebounding shock reaches the shell boundary at $t\sim1.2$ ns. Because of fuel-shell diffusion, this boundary is not sharp. We choose two contours corresponding to molar shell SiO$_2$ ion concentrations of 10$\%$ and 90$\%$ to mark the fuel-shell interface [the magenta countours in Figs.~\ref{fig:burn}(a-d)], and we refer to the region between these two contours as the "mix layer." For comparison, the fuel-shell interface from the corresponding HYDRA simulations are plotted (dashed green curves), and the peak compression in HYDRA is greater. The capsule size inferred from X ray images of these experiments is larger than predicted by HYDRA \cite{herrmann:2009} and closer to the hot spot size in the LSP calculations.  

The calculated rate, $N$, of DT fusion neutrons produced per $\mu$m spherical shell is plotted in Figs.~\ref{fig:burn}(c-d), where it is defined as $N = n_D n_T <\sigma v> 4\pi r^2 * (1$  $\mu$m) with $<\sigma v>$ based on the effective temperature $T_{eff} = (m_TT_D+m_DT_T)/(m_D+m_T)$ \cite{bellei:2013}. Here, temperatures are defined by $T = (m/3n) \int |{\bf{v-u}}|^2 f({\bf{v}})d^3v$, where $\bf{u}$ is the mean flow of the species. The net burn histories are plotted in Figs.~\ref{fig:burn}(e-f). Because these two experiments are not hydro-equivalent, the differences between their burn profiles are not attributable to multi-species or kinetic effects on their own. The observational data comes from a gas Cerenkov detector sensitive to DT $\gamma$ production, with an instrumental uncertainty of $\sim70$ ps. The LSP burn history [in black in Fig.~\ref{fig:burn}(e-f)] is derived from a Monte Carlo binary fusion model, and it may be compared to the integrated burn derived from $<\sigma v>$ (in blue).  The small discrepancy is possibly caused by non-Maxwellian ion velocity distributions \cite{larroche:2012} or tail depletion \cite{petschek:1979}. For our LSP calculations, the ion distributions are resolved out to $\sim$ 5---6 $T_i$. Over this range, we do not observe deviations from Maxwellian distributions as large as predicted by tail depletion models \cite{molvig:2012,albright:2013,kagan:2015}, possibly because the boundary conditions and other assumptions required to obtain analytic estimates are not well-satisfied in our calculations. 

Because the integrated burn in LSP is close to that inferred from a Maxwellian distribution, the discrepancy between LSP and HYDRA is mainly caused by differences in the density and temperature profiles. The burn-weighted ion temperatures $<T_i>$ inferred from neutron spectra \cite{brysk:1973,murphy:2014} in LSP, HYDRA, and experiment are nevertheless similar. They are, respectively, $<T_i>\sim$ 4.9, 5.2, and 5.1 keV for the D/T shot, and 4.7, 3.9, and 4.9 keV for the D/T/$^3$He shot. In agreement with experimental conclusions \cite{herrmann:2009}, the 50/50 D/T shot fusion yield is compression-dominated, and the 32/32/36 D/T/$^3$He case (which, again, is not hydro-equivalent) shock-dominated. The total yield based on local reactivity is $\sim5\%$ larger than the binary fusion model. The observed bang times for peak fusion rate are 1.40 ns (D/T)  and 1.28 ns (D/T/$^3$He), with $\sim50$ ps experimental resolution. The simulated bang times are 1.35 ns and 1.38 ns in HYDRA and 1.28 ns and 1.19 ns in LSP, though these vary by $\sim 100$ ps from the lowest laser factor considered to the highest. The simulated burn widths are 50 to 100 ps longer than observed. 

\begin{figure}
\includegraphics[width = 8.0cm]{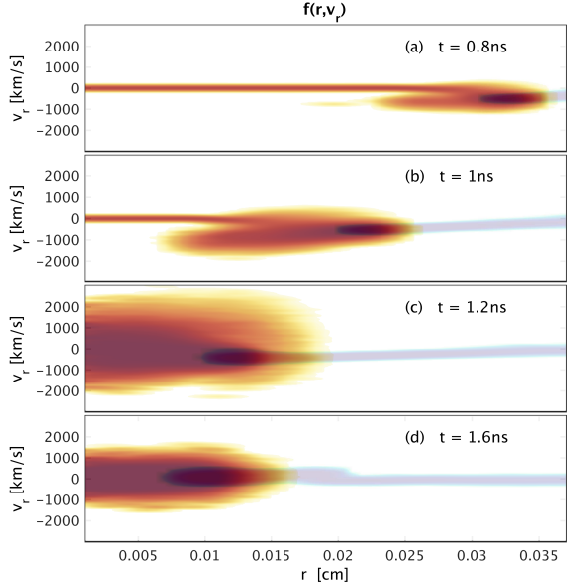}
\caption{Reduced phase space density $f(r,v_r)$ of D (red) and SiO$_2$ average (blue) ions at four time slices from a simulation of a direct-drive implosion of a D/T/$^3$He-filled glass shell. At the incoming shock front in (a) and (b), the D fuel ion distribution shows two well-separated populations, composed of shocked accelerated ions and cold unshocked fuel. Near shock convergence in (c), a non-Maxwellian tail of reflected ions is produced. Finally, past bang time (d), the ions have collisionally relaxed to a nearly Maxwellian velocity distribution. These qualitative features are similar for each fuel ion species. 
\label{fig:phase}}
\end{figure}

The reduced radial phase space density $f(r,v_r)$ of the D ions is plotted at four times in Fig.~\ref{fig:phase} from the D/T/$^3$He shot. Several qualitative features are the same for each ion species. The most kinetic regime is during shock convergence when the density is still relatively low, and a typical thermal fuel ion mean-free path $\lambda \gtrsim 100$ $\mu$m in the hot regions. During this phase, ion velocity distributions are highly non-Maxwellian. There are two well-separated populations in velocity space [Figs.~\ref{fig:phase}(a-b)] as shock-accelerated ions move to the forefront and mix with the cooler gas. Shortly after shock convergence in Fig.~\ref{fig:phase}(c), there remains a superthermal tail of energetic reflected ions moving back towards the shell. During the compression phase of the implosion, the density and collisionality are higher with $\lambda$ on the order of a few $\mu$m, and the ion distribution relaxes close to a Maxwellian throughout the domain. The Knudsen number \cite{rosenberg:2014}, defined as a ratio of $\lambda$ to a nominal system size $L$, varies significantly over the course of the implosion. For average values relevant to the fusion burn, we use reactivity-weighted quantities defined as follows. For any quantity $Q(r,t)$, the reactivity-weighted average $\overline{Q} = (1/Y)\int\int Q(r,t) R(r,t) 4\pi r^2 dr dt$, where $R(r,t)$ is the local Maxwellian fusion reactivity and $Y = \int\int R(r,t) 4\pi r^2 dr dt$ is the fusion yield. Reactivity-weighted mean-free paths based on density and temperature  profiles are in the range $\overline{\lambda_{D,T}}\sim$ 10---20 $\mu$m and $\overline{\lambda_{^3He}}\sim 5$ $\mu$m. For a hot spot size of $L\sim70$ $\mu$m, typical Knudsen numbers are $K<0.3$, implying these shots are in a moderately kinetic regime.

\begin{figure}
\includegraphics[width = 8.0cm]{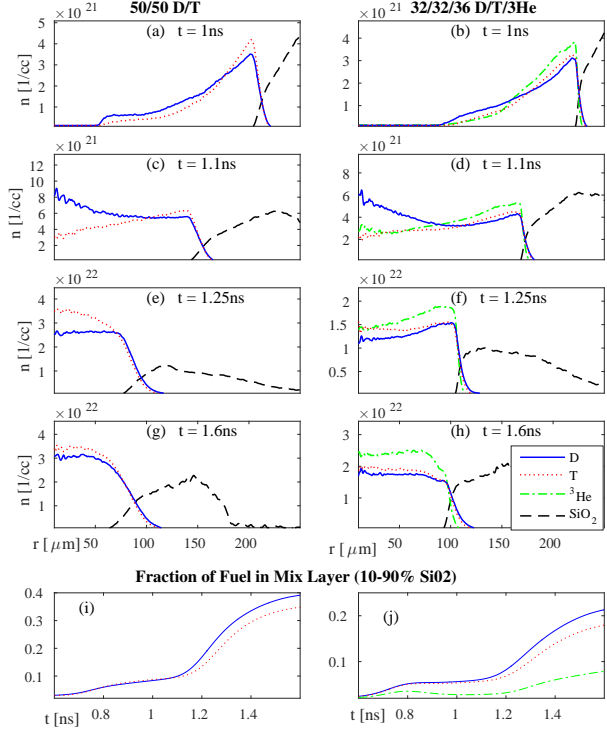}
\caption{Density profiles at four times from simulations of (a,c,e,g) a D/T-filled capsule and (b,d,f,h) a D/T/$^3$He-filled capsule. In each case, lighter D ion concentration is enhanced on the leading side of the shocked plasma (a-d). At later times (e-h), the hot spot becomes slightly richer in the heavier species. (i-j) Fraction of fuel ions of each species within the fuel-shell mix layer over time.
\label{fig:density}}
\end{figure}

Although the velocity space structure of each ion species is qualitatively similar, the bulk density and temperature profiles separate over time. In line with previous numerical studies \cite{bellei:2014} and transport theory \cite{kagan:2014}, the lighter ions move ahead of the heavier species across the incoming shock. The ion density profiles are plotted in Fig.~\ref{fig:density}. The largest change in relative ion concentrations occurs near the origin at shock convergence at $t\sim1.1$ ns [see Figs.~\ref{fig:density}(c-d)]. As the shock rebounds, the D ions again move fastest, leaving behind a T- and $^3$He-rich core. While the local ratio $n_D/n_T$ peaks at $\sim 2$ at shock convergence, this large stratification is limited to a small volume and a short time, similar to previous kinetic simulations \cite{larroche:2012}. The reactivity-weighted concentration ratios are $\overline{n_T}/\overline{n_D}=1.08$ for 50/50 D/T, and $\overline{n_T}/\overline{n_D}=1.04$ and $\overline{n_{^3He}}/\overline{n_D}=1.25$ for 32/32/36 D/T/$^3$He (with initially $n_{^3He}/n_D=1.125$). In a single-fluid description with a 50/50 D/T fuel mix, the total yield is the reactivity $R\sim(1/4) n^2<\sigma v>$ integrated over the fuel volume. As an upper bound on the effect of density stratification on fusion yield in these simulations, we find the yields differ at most $\sim12\%$ when substituting $n=2*n_D$ or $n=2*n_T$ for total fuel density in $R$. 

The density profiles also indicate that the fraction of fuel within the mix layer [plotted in Figs.~\ref{fig:density}(i-j)] increases over time. A crude estimate of the mixed fuel fraction is obtained by finding the typical length scale $\Delta R$ of the mix layer based on a diffusion coefficient $D\sim \nu_{FS} \lambda_F^2$, where $\nu_{FS}$ ($F$uel with $S$hell) is a collision frequency and $\lambda_F\sim \sqrt{2T_F/m_F}/\nu_{FS}$ is a fuel mean-free path. The characteristic mix length is then $\Delta R \sim \sqrt{2\tau D}$, where $\tau$ is a typical time-scale for the diffusion. We have
\begin{equation}
\Delta R \sim (14 \mbox{$\mu$m})*\sqrt{\tau_{ns}\frac{T_{keV}^{5/2}}{n_{23}}\frac{1}{\mu_F^{1/2}Z_F^2 Z_S^2}},
\nonumber
\end{equation}
where $\tau_{ns}$ is in nanoseconds, $T_{keV}$ is the ion temperature in keV within the mix layer, $n_{23}$ is the local shell density in units of $10^{23}$/cc, and $\mu_F$ is the fuel ion mass divided by the proton mass, and we take a Coulomb logarithm $\ln\Lambda\sim9$. In our cases, typical values are $\tau_{ns}\sim0.5$, $T_{keV}\sim2.5$, $n_{23}\sim0.2$, and $Z_S=10$, yielding a mixing length of $\Delta R\sim 6$ $\mu$m for the D ions. Note that the full width of the mix layer, defined by 10$\%$ and 90$\%$ SiO$_2$ contours in Figs.~\ref{fig:burn}(a-d), depends on the shape and tail of the density profile and is typically 4---5 $\Delta R$. Assuming the fuel density profile is uniform out to the mix layer at radius $R$, the approximate mixed fuel fraction is then $f\sim 3\Delta R/R$ \cite{molvig:2014}. Our estimate yields mixed fuel fractions of $f\sim20\%$, in rough agreement with Figs.~\ref{fig:density}(i-j). Because it is cooled and diluted, the mixed fuel only contributes $7\%$ and $2\%$ of the total fusion yield for the D/T and D/T/$^3$He shots. Furthermore, assuming the fusion yield scales as $\propto n^2$, the density diffusion alone could reduce the total yield $\sim 30\%$. Note that the $^3$He ions with $Z=2$ diffuse less, and a higher concentration of $^3$He persists throughout the hot spot [see Fig.\ref{fig:density}(h)].

\begin{figure}
\includegraphics[width = 8.0cm]{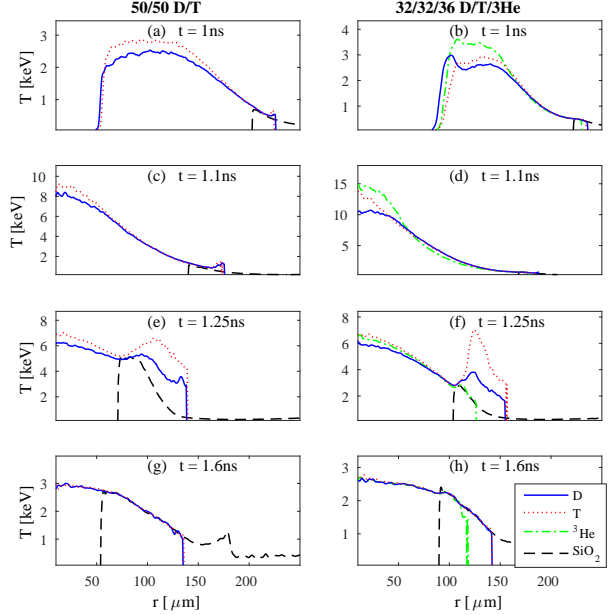}
\caption{Temperature profiles at four times from simulations of (a,c,e,g) a D/T-filled capsule and (b,d,f,h) a D/T/He3-filled capsule. The heavier species T and $^3$He are heated to higher temperatures than the lighter species D by the shock. Shortly after shock convergence (e-f), a small number of high-energy tail fuel ions penetrate into the shell and raise the effective fuel temperature in the mix layer. At later times (g-h), the temperatures have equilibrated.
\label{fig:temp}}
\end{figure}

The ion temperatures also separate during the implosion, as plotted in Fig.~\ref{fig:temp}. In the converging shock, the heavier ion species reach higher peak temperatures, as in previous studies \cite{larroche:2016,bellei:2014}. Immediately following shock convergence, a relatively small number of energetic reflected ions [see Fig.~\ref{fig:phase}(b)] escape into the fuel-shell mix layer and form a peak in the temperature [Figs.~\ref{fig:temp}(e-f)]. They are rapidly slowed down, however, and they equilibrate with the shell ions within tens of picoseconds. At $t=1.3$ ns, the fuel ions within the mix layer carry $\sim10\%$ of the total thermal energy of the fuel, defined as a volume integral of $(3/2)(n_DT_D+n_TT_T)$. In addition, we estimate that $\sim10\%$ of the total fuel thermal energy has been transferred to the heated SiO$_2$ ions in the mix layer, where the estimated transferred energy is $(3/2)n_{S}(T_{S}-T_0)$ integrated over the mix layer, with $T_0\sim$ 500 eV the typical shell temperature immediately outside the mix layer. As with the density separation, the temperature separation within the hot spot is more significant during the shock convergence phase. In terms of Maxwellian reactivity-weighted averages, we find $\overline{T_T}/\overline{T_D}=1.07$ (for 50/50 D/T), and $\overline{T_T}/\overline{T_D}=1.08$ and $\overline{T_{^3He}}/\overline{T_D}=1.06$ (for 32/32/36 D/T/$^3$He). The total computed fusion yield would increase by $\sim20\%$ with $<\sigma v>$ based on $T_T$ compared to $T_D$.

\section{Discussion and Summary}

It is important to question how the diffusive and kinetic effects we examined change with deviations from spherical symmetry. The density separation near the origin at shock convergence, for example, will likely be less extreme in 3D systems where fronts from each direction need not converge simultaneously or precisely at the origin. As we recall, our estimate for the diffusion of fuel into the shell scales as $(S/V)\Delta R$, where $S/V$ is the surface-to-volume ratio of the fuel. This is minimized in a perfectly spherical geometry and could be significantly enhanced by hydrodynamic perturbations of the fuel-shell interface \cite{molvig:2014,kagan:2015}.

In conclusion, we modeled a pair of capsule implosion experiments using a hybrid kinetic model with a kinetic treatment of both the fuel and shell ions. Diffusion of fuel ions and energy into the shell played a large role in degrading the fusion yield. The density and temperature profiles of the ion species separated as the initial shock converged, but as in previous kinetic studies \cite{larroche:2012}, the species stratification was mild in the later compressive phase of the implosion. Shock-driven species separation likely occurs to some degree in ignition-relevant experiments, and it may alter the shock fusion yield. It remains unclear whether these early species separation effects could leave a lasting mark on the fusion performance of ignition capsules, while there is some numerical evidence that both fuel \cite{inglebert:2014} and fusion product \cite{peigney:2014} kinetics may degrade ignition capsule performance. As hybrid kinetic modeling \cite{welch:2001,larroche:2012,taitano:2015} becomes more developed, it could help further explore kinetic effects under ignition conditions.

\begin{acknowledgments}
We would like to thank C.~Bellei and S.~Wilks for sharing the input deck of their LSP simulation and for useful advice and E.~Dodd for helpful discussion and computing support. D.~Welch and C.~Thoma of the LSP team at Voss Scientific deserve our special thanks for their invaluable guidance. This work was performed under the auspices of Los Alamos National Laboratory under US DOE/NNSA contract DE-AC52-06NA25396. Simulations were performed with LANL IC and ASC resources. 
\end{acknowledgments}


\begin{thebibliography}{32}
\expandafter\ifx\csname natexlab\endcsname\relax\def\natexlab#1{#1}\fi
\expandafter\ifx\csname bibnamefont\endcsname\relax
  \def\bibnamefont#1{#1}\fi
\expandafter\ifx\csname bibfnamefont\endcsname\relax
  \def\bibfnamefont#1{#1}\fi
\expandafter\ifx\csname citenamefont\endcsname\relax
  \def\citenamefont#1{#1}\fi
\expandafter\ifx\csname url\endcsname\relax
  \def\url#1{\texttt{#1}}\fi
\expandafter\ifx\csname urlprefix\endcsname\relax\def\urlprefix{URL }\fi
\providecommand{\bibinfo}[2]{#2}
\providecommand{\eprint}[2][]{\url{#2}}

\bibitem[{\citenamefont{Rygg et~al.}(2006)\citenamefont{Rygg, Frenje, Li,
  S{\'e}guin, Petrasso, Delettrez, Glebov, Goncharov, Meyerhofer, Regan
  et~al.}}]{rygg:2006}
\bibinfo{author}{\bibfnamefont{J.~R.} \bibnamefont{Rygg}},
  \bibinfo{author}{\bibfnamefont{J.~A.} \bibnamefont{Frenje}},
  \bibinfo{author}{\bibfnamefont{C.-K.} \bibnamefont{Li}},
  \bibinfo{author}{\bibfnamefont{F.~H.} \bibnamefont{S{\'e}guin}},
  \bibinfo{author}{\bibfnamefont{R.}~\bibnamefont{Petrasso}},
  \bibinfo{author}{\bibfnamefont{J.~A.} \bibnamefont{Delettrez}},
  \bibinfo{author}{\bibfnamefont{V.~Y.} \bibnamefont{Glebov}},
  \bibinfo{author}{\bibfnamefont{V.~N.} \bibnamefont{Goncharov}},
  \bibinfo{author}{\bibfnamefont{D.~D.} \bibnamefont{Meyerhofer}},
  \bibinfo{author}{\bibfnamefont{S.~P.} \bibnamefont{Regan}},
  \bibnamefont{et~al.}, \bibinfo{journal}{Physics of Plasmas (1994-present)}
  \textbf{\bibinfo{volume}{13}}, \bibinfo{pages}{052702}
  (\bibinfo{year}{2006}).

\bibitem[{\citenamefont{Herrmann et~al.}(2009)\citenamefont{Herrmann,
  Langenbrunner, Mack, Cooley, Wilson, Evans, Sedillo, Kyrala, Caldwell, Young
  et~al.}}]{herrmann:2009}
\bibinfo{author}{\bibfnamefont{H.}~\bibnamefont{Herrmann}},
  \bibinfo{author}{\bibfnamefont{J.}~\bibnamefont{Langenbrunner}},
  \bibinfo{author}{\bibfnamefont{J.}~\bibnamefont{Mack}},
  \bibinfo{author}{\bibfnamefont{J.}~\bibnamefont{Cooley}},
  \bibinfo{author}{\bibfnamefont{D.}~\bibnamefont{Wilson}},
  \bibinfo{author}{\bibfnamefont{S.}~\bibnamefont{Evans}},
  \bibinfo{author}{\bibfnamefont{T.}~\bibnamefont{Sedillo}},
  \bibinfo{author}{\bibfnamefont{G.}~\bibnamefont{Kyrala}},
  \bibinfo{author}{\bibfnamefont{S.}~\bibnamefont{Caldwell}},
  \bibinfo{author}{\bibfnamefont{C.}~\bibnamefont{Young}},
  \bibnamefont{et~al.}, \bibinfo{journal}{Physics of Plasmas (1994-present)}
  \textbf{\bibinfo{volume}{16}}, \bibinfo{pages}{056312}
  (\bibinfo{year}{2009}).

\bibitem[{\citenamefont{Casey et~al.}(2012)\citenamefont{Casey, Frenje,
  Gatu~Johnson, Manuel, Rinderknecht, Sinenian, S\'eguin, Li, Petrasso, Radha
  et~al.}}]{casey:2012}
\bibinfo{author}{\bibfnamefont{D.~T.} \bibnamefont{Casey}},
  \bibinfo{author}{\bibfnamefont{J.~A.} \bibnamefont{Frenje}},
  \bibinfo{author}{\bibfnamefont{M.}~\bibnamefont{Gatu~Johnson}},
  \bibinfo{author}{\bibfnamefont{M.~J.-E.} \bibnamefont{Manuel}},
  \bibinfo{author}{\bibfnamefont{H.~G.} \bibnamefont{Rinderknecht}},
  \bibinfo{author}{\bibfnamefont{N.}~\bibnamefont{Sinenian}},
  \bibinfo{author}{\bibfnamefont{F.~H.} \bibnamefont{S\'eguin}},
  \bibinfo{author}{\bibfnamefont{C.~K.} \bibnamefont{Li}},
  \bibinfo{author}{\bibfnamefont{R.~D.} \bibnamefont{Petrasso}},
  \bibinfo{author}{\bibfnamefont{P.~B.} \bibnamefont{Radha}},
  \bibnamefont{et~al.}, \bibinfo{journal}{Phys. Rev. Lett.}
  \textbf{\bibinfo{volume}{108}}, \bibinfo{pages}{075002}
  (\bibinfo{year}{2012}),
  \urlprefix\url{http://link.aps.org/doi/10.1103/PhysRevLett.108.075002}.

\bibitem[{\citenamefont{Rosenberg et~al.}(2014)\citenamefont{Rosenberg,
  Zylstra, S{\'e}guin, Rinderknecht, Frenje, Johnson, Sio, Waugh, Sinenian, Li
  et~al.}}]{rosenberg:2014}
\bibinfo{author}{\bibfnamefont{M.}~\bibnamefont{Rosenberg}},
  \bibinfo{author}{\bibfnamefont{A.}~\bibnamefont{Zylstra}},
  \bibinfo{author}{\bibfnamefont{F.}~\bibnamefont{S{\'e}guin}},
  \bibinfo{author}{\bibfnamefont{H.}~\bibnamefont{Rinderknecht}},
  \bibinfo{author}{\bibfnamefont{J.}~\bibnamefont{Frenje}},
  \bibinfo{author}{\bibfnamefont{M.~G.} \bibnamefont{Johnson}},
  \bibinfo{author}{\bibfnamefont{H.}~\bibnamefont{Sio}},
  \bibinfo{author}{\bibfnamefont{C.}~\bibnamefont{Waugh}},
  \bibinfo{author}{\bibfnamefont{N.}~\bibnamefont{Sinenian}},
  \bibinfo{author}{\bibfnamefont{C.}~\bibnamefont{Li}}, \bibnamefont{et~al.},
  \bibinfo{journal}{Physics of Plasmas (1994-present)}
  \textbf{\bibinfo{volume}{21}}, \bibinfo{pages}{122712}
  (\bibinfo{year}{2014}).

\bibitem[{\citenamefont{Rinderknecht et~al.}(2015)\citenamefont{Rinderknecht,
  Rosenberg, Li, Hoffman, Kagan, Zylstra, Sio, Frenje, Gatu~Johnson, S\'eguin
  et~al.}}]{rinderknecht:2015}
\bibinfo{author}{\bibfnamefont{H.~G.} \bibnamefont{Rinderknecht}},
  \bibinfo{author}{\bibfnamefont{M.~J.} \bibnamefont{Rosenberg}},
  \bibinfo{author}{\bibfnamefont{C.~K.} \bibnamefont{Li}},
  \bibinfo{author}{\bibfnamefont{N.~M.} \bibnamefont{Hoffman}},
  \bibinfo{author}{\bibfnamefont{G.}~\bibnamefont{Kagan}},
  \bibinfo{author}{\bibfnamefont{A.~B.} \bibnamefont{Zylstra}},
  \bibinfo{author}{\bibfnamefont{H.}~\bibnamefont{Sio}},
  \bibinfo{author}{\bibfnamefont{J.~A.} \bibnamefont{Frenje}},
  \bibinfo{author}{\bibfnamefont{M.}~\bibnamefont{Gatu~Johnson}},
  \bibinfo{author}{\bibfnamefont{F.~H.} \bibnamefont{S\'eguin}},
  \bibnamefont{et~al.}, \bibinfo{journal}{Phys. Rev. Lett.}
  \textbf{\bibinfo{volume}{114}}, \bibinfo{pages}{025001}
  (\bibinfo{year}{2015}),
  \urlprefix\url{http://link.aps.org/doi/10.1103/PhysRevLett.114.025001}.

\bibitem[{\citenamefont{Amendt et~al.}(2010)\citenamefont{Amendt, Landen,
  Robey, Li, and Petrasso}}]{amendt:2010}
\bibinfo{author}{\bibfnamefont{P.}~\bibnamefont{Amendt}},
  \bibinfo{author}{\bibfnamefont{O.}~\bibnamefont{Landen}},
  \bibinfo{author}{\bibfnamefont{H.}~\bibnamefont{Robey}},
  \bibinfo{author}{\bibfnamefont{C.}~\bibnamefont{Li}}, \bibnamefont{and}
  \bibinfo{author}{\bibfnamefont{R.}~\bibnamefont{Petrasso}},
  \bibinfo{journal}{Physical review letters} \textbf{\bibinfo{volume}{105}},
  \bibinfo{pages}{115005} (\bibinfo{year}{2010}).

\bibitem[{\citenamefont{Amendt et~al.}(2011)\citenamefont{Amendt, Wilks,
  Bellei, Li, and Petrasso}}]{amendt:2011}
\bibinfo{author}{\bibfnamefont{P.}~\bibnamefont{Amendt}},
  \bibinfo{author}{\bibfnamefont{S.}~\bibnamefont{Wilks}},
  \bibinfo{author}{\bibfnamefont{C.}~\bibnamefont{Bellei}},
  \bibinfo{author}{\bibfnamefont{C.}~\bibnamefont{Li}}, \bibnamefont{and}
  \bibinfo{author}{\bibfnamefont{R.}~\bibnamefont{Petrasso}},
  \bibinfo{journal}{Physics of Plasmas (1994-present)}
  \textbf{\bibinfo{volume}{18}}, \bibinfo{pages}{056308}
  (\bibinfo{year}{2011}).

\bibitem[{\citenamefont{Amendt et~al.}(2015)\citenamefont{Amendt, Bellei, Ross,
  and Salmonson}}]{amendt:2015}
\bibinfo{author}{\bibfnamefont{P.}~\bibnamefont{Amendt}},
  \bibinfo{author}{\bibfnamefont{C.}~\bibnamefont{Bellei}},
  \bibinfo{author}{\bibfnamefont{J.~S.} \bibnamefont{Ross}}, \bibnamefont{and}
  \bibinfo{author}{\bibfnamefont{J.}~\bibnamefont{Salmonson}},
  \bibinfo{journal}{Physical Review E} \textbf{\bibinfo{volume}{91}},
  \bibinfo{pages}{023103} (\bibinfo{year}{2015}).

\bibitem[{\citenamefont{Bellei et~al.}(2013)\citenamefont{Bellei, Amendt,
  Wilks, Haines, Casey, Li, Petrasso, and Welch}}]{bellei:2013}
\bibinfo{author}{\bibfnamefont{C.}~\bibnamefont{Bellei}},
  \bibinfo{author}{\bibfnamefont{P.~A.} \bibnamefont{Amendt}},
  \bibinfo{author}{\bibfnamefont{S.~C.} \bibnamefont{Wilks}},
  \bibinfo{author}{\bibfnamefont{M.~G.} \bibnamefont{Haines}},
  \bibinfo{author}{\bibfnamefont{D.~T.} \bibnamefont{Casey}},
  \bibinfo{author}{\bibfnamefont{C.~K.} \bibnamefont{Li}},
  \bibinfo{author}{\bibfnamefont{R.}~\bibnamefont{Petrasso}}, \bibnamefont{and}
  \bibinfo{author}{\bibfnamefont{D.~R.} \bibnamefont{Welch}},
  \bibinfo{journal}{Physics of Plasmas} \textbf{\bibinfo{volume}{20}},
  \bibinfo{eid}{012701} (\bibinfo{year}{2013}),
  \urlprefix\url{http://scitation.aip.org/content/aip/journal/pop/20/1/10.1063%
/1.4773291}.

\bibitem[{\citenamefont{Bellei and Amendt}(2014)}]{bellei:2014}
\bibinfo{author}{\bibfnamefont{C.}~\bibnamefont{Bellei}} \bibnamefont{and}
  \bibinfo{author}{\bibfnamefont{P.~A.} \bibnamefont{Amendt}},
  \bibinfo{journal}{Phys. Rev. E} \textbf{\bibinfo{volume}{90}},
  \bibinfo{pages}{013101} (\bibinfo{year}{2014}),
  \urlprefix\url{http://link.aps.org/doi/10.1103/PhysRevE.90.013101}.

\bibitem[{\citenamefont{Larroche}(2012)}]{larroche:2012}
\bibinfo{author}{\bibfnamefont{O.}~\bibnamefont{Larroche}},
  \bibinfo{journal}{Physics of Plasmas (1994-present)}
  \textbf{\bibinfo{volume}{19}}, \bibinfo{pages}{122706}
  (\bibinfo{year}{2012}).

\bibitem[{\citenamefont{Larroche et~al.}(2016)\citenamefont{Larroche,
  Rinderknecht, Rosenberg, Hoffman, Atzeni, Petrasso, Amendt, and
  S{\'e}guin}}]{larroche:2016}
\bibinfo{author}{\bibfnamefont{O.}~\bibnamefont{Larroche}},
  \bibinfo{author}{\bibfnamefont{H.}~\bibnamefont{Rinderknecht}},
  \bibinfo{author}{\bibfnamefont{M.}~\bibnamefont{Rosenberg}},
  \bibinfo{author}{\bibfnamefont{N.}~\bibnamefont{Hoffman}},
  \bibinfo{author}{\bibfnamefont{S.}~\bibnamefont{Atzeni}},
  \bibinfo{author}{\bibfnamefont{R.}~\bibnamefont{Petrasso}},
  \bibinfo{author}{\bibfnamefont{P.}~\bibnamefont{Amendt}}, \bibnamefont{and}
  \bibinfo{author}{\bibfnamefont{F.}~\bibnamefont{S{\'e}guin}},
  \bibinfo{journal}{Physics of Plasmas (1994-present)}
  \textbf{\bibinfo{volume}{23}}, \bibinfo{pages}{012701}
  (\bibinfo{year}{2016}).

\bibitem[{\citenamefont{Marinak et~al.}(1998)\citenamefont{Marinak, Haan,
  Dittrich, Tipton, and Zimmerman}}]{marinak:1998}
\bibinfo{author}{\bibfnamefont{M.}~\bibnamefont{Marinak}},
  \bibinfo{author}{\bibfnamefont{S.}~\bibnamefont{Haan}},
  \bibinfo{author}{\bibfnamefont{T.}~\bibnamefont{Dittrich}},
  \bibinfo{author}{\bibfnamefont{R.}~\bibnamefont{Tipton}}, \bibnamefont{and}
  \bibinfo{author}{\bibfnamefont{G.}~\bibnamefont{Zimmerman}},
  \bibinfo{journal}{Physics of Plasmas (1994-present)}
  \textbf{\bibinfo{volume}{5}}, \bibinfo{pages}{1125} (\bibinfo{year}{1998}).

\bibitem[{\citenamefont{Welch et~al.}(2001)\citenamefont{Welch, Rose, Oliver,
  and Clark}}]{welch:2001}
\bibinfo{author}{\bibfnamefont{D.~R.} \bibnamefont{Welch}},
  \bibinfo{author}{\bibfnamefont{D.}~\bibnamefont{Rose}},
  \bibinfo{author}{\bibfnamefont{B.}~\bibnamefont{Oliver}}, \bibnamefont{and}
  \bibinfo{author}{\bibfnamefont{R.}~\bibnamefont{Clark}},
  \bibinfo{journal}{Nuclear Instruments and Methods in Physics Research Section
  A: Accelerators, Spectrometers, Detectors and Associated Equipment}
  \textbf{\bibinfo{volume}{464}}, \bibinfo{pages}{134} (\bibinfo{year}{2001}).

\bibitem[{\citenamefont{Welch et~al.}(2006)\citenamefont{Welch, Rose, Cuneo,
  Campbell, and Mehlhorn}}]{welch:2006}
\bibinfo{author}{\bibfnamefont{D.}~\bibnamefont{Welch}},
  \bibinfo{author}{\bibfnamefont{D.}~\bibnamefont{Rose}},
  \bibinfo{author}{\bibfnamefont{M.}~\bibnamefont{Cuneo}},
  \bibinfo{author}{\bibfnamefont{R.}~\bibnamefont{Campbell}}, \bibnamefont{and}
  \bibinfo{author}{\bibfnamefont{T.}~\bibnamefont{Mehlhorn}},
  \bibinfo{journal}{Physics of Plasmas (1994-present)}
  \textbf{\bibinfo{volume}{13}}, \bibinfo{pages}{063105}
  (\bibinfo{year}{2006}).

\bibitem[{\citenamefont{Thoma et~al.}(2011)\citenamefont{Thoma, Welch, Clark,
  Bruner, MacFarlane, and Golovkin}}]{thoma:2011}
\bibinfo{author}{\bibfnamefont{C.}~\bibnamefont{Thoma}},
  \bibinfo{author}{\bibfnamefont{D.}~\bibnamefont{Welch}},
  \bibinfo{author}{\bibfnamefont{R.}~\bibnamefont{Clark}},
  \bibinfo{author}{\bibfnamefont{N.}~\bibnamefont{Bruner}},
  \bibinfo{author}{\bibfnamefont{J.}~\bibnamefont{MacFarlane}},
  \bibnamefont{and} \bibinfo{author}{\bibfnamefont{I.}~\bibnamefont{Golovkin}},
  \bibinfo{journal}{Physics of Plasmas (1994-present)}
  \textbf{\bibinfo{volume}{18}}, \bibinfo{pages}{103507}
  (\bibinfo{year}{2011}).

\bibitem[{\citenamefont{Nanbu}(1997)}]{nanbu:1997}
\bibinfo{author}{\bibfnamefont{K.}~\bibnamefont{Nanbu}},
  \bibinfo{journal}{Physical Review E} \textbf{\bibinfo{volume}{55}},
  \bibinfo{pages}{4642} (\bibinfo{year}{1997}).

\bibitem[{\citenamefont{Takizuka and Abe}(1977)}]{takizuka:1977}
\bibinfo{author}{\bibfnamefont{T.}~\bibnamefont{Takizuka}} \bibnamefont{and}
  \bibinfo{author}{\bibfnamefont{H.}~\bibnamefont{Abe}},
  \bibinfo{journal}{Journal of Computational Physics}
  \textbf{\bibinfo{volume}{25}}, \bibinfo{pages}{205} (\bibinfo{year}{1977}).

\bibitem[{\citenamefont{Hewett and Langdon}(1987)}]{hewett:1987}
\bibinfo{author}{\bibfnamefont{D.~W.} \bibnamefont{Hewett}} \bibnamefont{and}
  \bibinfo{author}{\bibfnamefont{A.~B.} \bibnamefont{Langdon}},
  \bibinfo{journal}{Journal of Computational Physics}
  \textbf{\bibinfo{volume}{72}}, \bibinfo{pages}{121} (\bibinfo{year}{1987}).

\bibitem[{\citenamefont{Smithe et~al.}(2009)\citenamefont{Smithe, Cary, and
  Carlsson}}]{smithe:2009}
\bibinfo{author}{\bibfnamefont{D.~N.} \bibnamefont{Smithe}},
  \bibinfo{author}{\bibfnamefont{J.~R.} \bibnamefont{Cary}}, \bibnamefont{and}
  \bibinfo{author}{\bibfnamefont{J.~A.} \bibnamefont{Carlsson}},
  \bibinfo{journal}{Journal of Computational Physics}
  \textbf{\bibinfo{volume}{228}}, \bibinfo{pages}{7289} (\bibinfo{year}{2009}).

\bibitem[{\citenamefont{Marinak et~al.}(2001)\citenamefont{Marinak, Kerbel,
  Gentile, Jones, Munro, Pollaine, Dittrich, and Haan}}]{marinak:2001}
\bibinfo{author}{\bibfnamefont{M.}~\bibnamefont{Marinak}},
  \bibinfo{author}{\bibfnamefont{G.}~\bibnamefont{Kerbel}},
  \bibinfo{author}{\bibfnamefont{N.}~\bibnamefont{Gentile}},
  \bibinfo{author}{\bibfnamefont{O.}~\bibnamefont{Jones}},
  \bibinfo{author}{\bibfnamefont{D.}~\bibnamefont{Munro}},
  \bibinfo{author}{\bibfnamefont{S.}~\bibnamefont{Pollaine}},
  \bibinfo{author}{\bibfnamefont{T.}~\bibnamefont{Dittrich}}, \bibnamefont{and}
  \bibinfo{author}{\bibfnamefont{S.}~\bibnamefont{Haan}},
  \bibinfo{journal}{Physics of Plasmas (1994-present)}
  \textbf{\bibinfo{volume}{8}}, \bibinfo{pages}{2275} (\bibinfo{year}{2001}).

\bibitem[{\citenamefont{Petschek and Henderson}(1979)}]{petschek:1979}
\bibinfo{author}{\bibfnamefont{A.}~\bibnamefont{Petschek}} \bibnamefont{and}
  \bibinfo{author}{\bibfnamefont{D.}~\bibnamefont{Henderson}},
  \bibinfo{journal}{Nuclear Fusion} \textbf{\bibinfo{volume}{19}},
  \bibinfo{pages}{1678} (\bibinfo{year}{1979}).

\bibitem[{\citenamefont{Molvig et~al.}(2012)\citenamefont{Molvig, Hoffman,
  Albright, Nelson, and Webster}}]{molvig:2012}
\bibinfo{author}{\bibfnamefont{K.}~\bibnamefont{Molvig}},
  \bibinfo{author}{\bibfnamefont{N.~M.} \bibnamefont{Hoffman}},
  \bibinfo{author}{\bibfnamefont{B.}~\bibnamefont{Albright}},
  \bibinfo{author}{\bibfnamefont{E.~M.} \bibnamefont{Nelson}},
  \bibnamefont{and} \bibinfo{author}{\bibfnamefont{R.~B.}
  \bibnamefont{Webster}}, \bibinfo{journal}{Physical review letters}
  \textbf{\bibinfo{volume}{109}}, \bibinfo{pages}{095001}
  (\bibinfo{year}{2012}).

\bibitem[{\citenamefont{Kagan et~al.}(2015)\citenamefont{Kagan, Svyatskiy,
  Rinderknecht, Rosenberg, Zylstra, Huang, and McDevitt}}]{kagan:2015}
\bibinfo{author}{\bibfnamefont{G.}~\bibnamefont{Kagan}},
  \bibinfo{author}{\bibfnamefont{D.}~\bibnamefont{Svyatskiy}},
  \bibinfo{author}{\bibfnamefont{H.~G.} \bibnamefont{Rinderknecht}},
  \bibinfo{author}{\bibfnamefont{M.~J.} \bibnamefont{Rosenberg}},
  \bibinfo{author}{\bibfnamefont{A.~B.} \bibnamefont{Zylstra}},
  \bibinfo{author}{\bibfnamefont{C.-K.} \bibnamefont{Huang}}, \bibnamefont{and}
  \bibinfo{author}{\bibfnamefont{C.~J.} \bibnamefont{McDevitt}},
  \bibinfo{journal}{Phys. Rev. Lett.} \textbf{\bibinfo{volume}{115}},
  \bibinfo{pages}{105002} (\bibinfo{year}{2015}),
  \urlprefix\url{http://link.aps.org/doi/10.1103/PhysRevLett.115.105002}.

\bibitem[{\citenamefont{Albright et~al.}(2013)\citenamefont{Albright, Molvig,
  Huang, Simakov, Dodd, Hoffman, Kagan, and Schmit}}]{albright:2013}
\bibinfo{author}{\bibfnamefont{B.~J.} \bibnamefont{Albright}},
  \bibinfo{author}{\bibfnamefont{K.}~\bibnamefont{Molvig}},
  \bibinfo{author}{\bibfnamefont{C.-K.} \bibnamefont{Huang}},
  \bibinfo{author}{\bibfnamefont{A.~N.} \bibnamefont{Simakov}},
  \bibinfo{author}{\bibfnamefont{E.~S.} \bibnamefont{Dodd}},
  \bibinfo{author}{\bibfnamefont{N.~M.} \bibnamefont{Hoffman}},
  \bibinfo{author}{\bibfnamefont{G.}~\bibnamefont{Kagan}}, \bibnamefont{and}
  \bibinfo{author}{\bibfnamefont{P.~F.} \bibnamefont{Schmit}},
  \bibinfo{journal}{Physics of Plasmas} \textbf{\bibinfo{volume}{20}},
  \bibinfo{eid}{122705} (\bibinfo{year}{2013}),
  \urlprefix\url{http://scitation.aip.org/content/aip/journal/pop/20/12/10.106%
3/1.4833639;jsessionid=BsfSQluQ05SMKiY8J35MbMPl.x-aip-live-02}.

\bibitem[{\citenamefont{Brysk}(1973)}]{brysk:1973}
\bibinfo{author}{\bibfnamefont{H.}~\bibnamefont{Brysk}},
  \bibinfo{journal}{Plasma Physics} \textbf{\bibinfo{volume}{15}},
  \bibinfo{pages}{611} (\bibinfo{year}{1973}).

\bibitem[{\citenamefont{Murphy}(2014)}]{murphy:2014}
\bibinfo{author}{\bibfnamefont{T.}~\bibnamefont{Murphy}},
  \bibinfo{journal}{Physics of Plasmas (1994-present)}
  \textbf{\bibinfo{volume}{21}}, \bibinfo{pages}{072701}
  (\bibinfo{year}{2014}).

\bibitem[{\citenamefont{Kagan and Tang}(2014)}]{kagan:2014}
\bibinfo{author}{\bibfnamefont{G.}~\bibnamefont{Kagan}} \bibnamefont{and}
  \bibinfo{author}{\bibfnamefont{X.-Z.} \bibnamefont{Tang}},
  \bibinfo{journal}{Physics Letters A} \textbf{\bibinfo{volume}{378}},
  \bibinfo{pages}{1531} (\bibinfo{year}{2014}).

\bibitem[{\citenamefont{Molvig et~al.}(2014)\citenamefont{Molvig, Vold, Dodd,
  and Wilks}}]{molvig:2014}
\bibinfo{author}{\bibfnamefont{K.}~\bibnamefont{Molvig}},
  \bibinfo{author}{\bibfnamefont{E.~L.} \bibnamefont{Vold}},
  \bibinfo{author}{\bibfnamefont{E.~S.} \bibnamefont{Dodd}}, \bibnamefont{and}
  \bibinfo{author}{\bibfnamefont{S.~C.} \bibnamefont{Wilks}},
  \bibinfo{journal}{Physical review letters} \textbf{\bibinfo{volume}{113}},
  \bibinfo{pages}{145001} (\bibinfo{year}{2014}).

\bibitem[{\citenamefont{Inglebert et~al.}(2014)\citenamefont{Inglebert, Canaud,
  and Larroche}}]{inglebert:2014}
\bibinfo{author}{\bibfnamefont{A.}~\bibnamefont{Inglebert}},
  \bibinfo{author}{\bibfnamefont{B.}~\bibnamefont{Canaud}}, \bibnamefont{and}
  \bibinfo{author}{\bibfnamefont{O.}~\bibnamefont{Larroche}},
  \bibinfo{journal}{EPL (Europhysics Letters)} \textbf{\bibinfo{volume}{107}},
  \bibinfo{pages}{65003} (\bibinfo{year}{2014}),
  \urlprefix\url{http://stacks.iop.org/0295-5075/107/i=6/a=65003}.

\bibitem[{\citenamefont{Peigney et~al.}(2014)\citenamefont{Peigney, Larroche,
  and Tikhonchuk}}]{peigney:2014}
\bibinfo{author}{\bibfnamefont{B.}~\bibnamefont{Peigney}},
  \bibinfo{author}{\bibfnamefont{O.}~\bibnamefont{Larroche}}, \bibnamefont{and}
  \bibinfo{author}{\bibfnamefont{V.}~\bibnamefont{Tikhonchuk}},
  \bibinfo{journal}{Physics of Plasmas (1994-present)}
  \textbf{\bibinfo{volume}{21}}, \bibinfo{pages}{122709}
  (\bibinfo{year}{2014}).

\bibitem[{\citenamefont{Taitano et~al.}(2015)\citenamefont{Taitano, Chac{\'o}n,
  Simakov, and Molvig}}]{taitano:2015}
\bibinfo{author}{\bibfnamefont{W.}~\bibnamefont{Taitano}},
  \bibinfo{author}{\bibfnamefont{L.}~\bibnamefont{Chac{\'o}n}},
  \bibinfo{author}{\bibfnamefont{A.}~\bibnamefont{Simakov}}, \bibnamefont{and}
  \bibinfo{author}{\bibfnamefont{K.}~\bibnamefont{Molvig}},
  \bibinfo{journal}{Journal of Computational Physics}
  \textbf{\bibinfo{volume}{297}}, \bibinfo{pages}{357} (\bibinfo{year}{2015}).

\end{thebibliography}


\end{document}